# Fractal continuum model for the adsorption-diffusion process

E. C. Herrera-Hernández[a*], C. G. Aguilar-Madera[b], R. Ocampo-Perez[c], G. Espinosa-Paredes[d], M. Núñez-López[e]

[a]*CONACYT-Centro de Ingeniería y Desarrollo Industrial, Querétaro, Qro. 76125, Mexico.*
[b]*Universidad Autónoma de Nuevo León, Facultad de Ciencias de la Tierra, Linares N.L, 67700, Mexico*
[c]*Facultad de Ciencias Químicas, Universidad Autónoma de San Luis Potosí, San Luis Potosí, 78260, Mexico*
[d]*Área de Ingeniería en Recursos Energéticos, Universidad Autónoma Metropolitana-Iztapalapa, Cd. México 09340, Mexico*
[e]*Department of Mathematics, ITAM, Río Hondo 1, Ciudad de México, 01080, Mexico*

**Abstract**

In this work, we present a mathematical model to describe the adsorption-diffusion process on fractal porous materials. This model is based on the fractal continuum approach and considers the scale-invariant properties of the surface and volume of adsorbent particles, which are well-represented by their fractal dimensions. The method of lines was used to solve the nonlinear fractal model, and the numerical predictions were compared with experimental data to determine the fractal dimensions through an optimization algorithm. The intraparticle mass flux and the mean square displacement dynamics as a function of fractal dimensions were analyzed. The results suggest that they can be potentially used to characterize the intraparticle mass transport processes. The fractal model demonstrated to be able to predict adsorption-diffusion experiments and jointly can be used to estimate fractal parameters of porous adsorbents.

*Corresponding author: *CONACYT-Centro de Ingeniería y Desarrollo Industrial, Querétaro*,

Av. Playa pie de la cuesta 702, Desarrollo San Pablo, Querétaro, Qro. 76125, Mexico,

(+52) 4425590385, erik.herrera@cidesi.edu.mx

**1. Introduction**

The adsorption-diffusion process is highly relevant in many physical, chemical and environmental natural and industrial processes (Dąbrowski, 2001). Several engineering problems such as pollutant removal (Xu et al., 2017, and Kyzas and Matis, 2015), heterogeneous catalysis (Oh et al. 2016 and Latimer et al. 2017), hydrogen storage and production (Dincer and Acar, 2015, Nikolaidis and Poullikkas, 2017, and Rezaei-Shouroki et al., 2017), driers and dehumidifiers (Burnett et al., 2015 and Bui et al., 201), and purification processes (Ardi et al., 2015, and Papathanasiou et al., 2016) involve adsorption-diffusion processes. As is well known, mathematical models, based on physical insights, are necessary to interpret the laboratory measurements obtained from controlled experiments or to describe the field data taken under real conditions. Concerning the adsorption process, a number of theoretical developments have been proposed after Langmuir's seminal work, which was formulated for gaining insights into the chemisorption (physisorption) process taking place during a chemical reaction on a hypothetical smooth (plane) solid surface (Langmuir, 1918). It is a fact that real adsorption surfaces are not smooth; therefore, more complex theories have been developed to address this issue. According to Avnir et al. (1984), at the molecular level, the surfaces of most material present self-similarity in a certain range of scales where the structure remains invariant. In the work of Pfeifer (1984), different methods to estimate the fractal dimension were proposed, and he found that the surface area scales as $A \sim r^{D-3}$, where $D$ represents the fractal dimension. Segars and Piscitelle (1996) used fractal theory to generalize the BET model by proposing a power-law parameter that quantifies the surface roughness. They interpret equilibrium data finding that the isothermic adsorption strongly depends on the fractal dimension, where $D > 2$. Following the same idea, the effect of fractal dimension on the adsorption process was analyzed, which allowed for modifications to the Brunauer–Emmett–Teller equation to improve predictions (Aguerre et al., 1996, Khalili et al., 1997, and Sun et al., 2015 ). Other work focused on improving the isotherm predictions was made by Kanô et al. (2000) who presented a model that generalizes both the Langmuir and Freundlich isotherms. Such a model is based on the fact that the adsorbing surface depends on the amount of adsorbed mass according to $A \sim M^\zeta$, where $\zeta = D/3$ quantifies the irregularities of a presumable self-similar fractal surface. Along the same line of ideas, in (Wang et al., 2007, Longjun et al., 2008, and Selmi et al., 2018), the authors present fractal models through the modification of the classical equilibrium isotherms. In such works, they discuss and apply a fractal adsorption model obtained from the Langmuir and kinetics model where the geometric characteristics of the solute are linked to the surface's fractal dimension. Additionally, they report a power-law dependence of the effective reaction order with time.

On the other hand, dynamic models to interpret the adsorption-diffusion process scarcely have changed from the classical descriptions commonly used for kinetic models (Qiu et al., 2009, Foo and Hameed, 2010, and Montagnaro and Balsamo, 2014) or diffusional models (Leyva-Ramos and Geankoplis 1994, Ocampo-Perez et al., 2010, Ocampo-Perez et al., 2013 and Ocampo-Pérez et al., 2017). Such approaches do not consider long-term correlations and long-range interactions appearing in diffusion processes occurring on complex domains such as self-similar (or fractal) structures. The work of Sakaguchi (2005) incorporates anomalous diffusion concepts into the analysis of adsorption-diffusion dynamics on fractal objects. Through direct simulations, Sakaguchi found that the adsorbed quantity followed a power-law time-dependent diffusion coefficient, i.e., $q(t) \sim t^{D_f/d_w}$. Conversely, Jiang et al. (2013) and Kang et al. (2015) predict methane adsorption/desorption data by using anomalous diffusion-based models that include temporal and spatial memory effects. They determine that this sort of model is better suited to describing experimental data than the classic Fikian-type ones (also known as normal diffusion). In the same manner, dos Santos *et al.* (2017) use a fractional adsorption-diffusion-based model to study the behavior of two chemical species being adsorbed on the same adsorbate but subjected to an irreversible reaction. Baumer and Stanger (2013) use the fractional fractal diffusion model coupled with the adsorption dynamic governed by the fractional equation to determine the transport behavior during the adsorption-diffusion process. They analyze and compare experimental data with theoretical predictions by changing the fractal parameters and the derivative order.

In this work, we developed a pore volume and surface diffusion model that incorporates anomalous diffusion features, with the difference from previous works being that our proposal is based on the fractal continuum approach that considers the fractality in the adsorbate geometrical characteristics. Specifically, we evaluate the adsorption-diffusion dynamics (solution concentration decay, intraparticle concentration profile, mass flux and square mean displacement) through a novel effective mathematical model that considers the accumulation rate within adsorbent particles due to the mass flux occurring not only on their surface but also because of the inner mass flux (pore volume diffusion). Such a model considers the fractality in the geometric aspects of adsorbent particles (surface and volume), and it uses fractal continuum theory to associate their fractal characteristics to an equivalent Euclidean representation without using more complex fractional operators, but rather the scaling factor instead.

## 2. Mathematical model formulation

Classical diffusion-adsorption models consider the homogeneous solute concentration in the aqueous solution since it depends on the amount of solute transported to the particles-solution interregion. Since mass transport inside the adsorbent material controls the global process, the solute concentration dynamic in the aqueous solution is governed by the following ordinary differential equation:

$$V \frac{dC_A}{dt} = -m\, S\, K_L \left(C_A - C_{r=R}\right) \qquad (1)$$

where $C_A$ and $C_{r=R}$ correspond to the solute concentration in the aqueous solution and at the solution-adsorbent interregion, respectively. Note that such a dynamic strongly depends on the driving force created by the difference between concentrations at the interface solution-particle. Furthermore, model parameters such as the solute mass ($m$), the contact interfacial area ($S$), the mass transfer coefficient ($K_L$) and the solution volume ($V$) uniquely determine the solute's dynamic behavior in the solution.

Since the proposed model contemplates adsorbent particles having the same absorption and geometrical properties, the mathematical expression we develop in this work considers the following assumptions.

- The adsorbent particles are of spherical geometry.
- The adsorption-diffusion phenomenon occurs in the same way in all adsorbent particles, and so it is enough to develop a model for one of them to describe the whole process.
- The adsorbent particles' size distribution is unimodal with a small variance.
- The particles present a self-similar structure in their surface and volume as the ratio changes. Such fractal characteristics depend on the fractal dimensions.
- By using the proper measures, it is possible to map the fractal surface and volume elements into their similar Euclidean ones. This mapping process provides the mathematical tools to represent a complex fractal structure in an equivalent homogeneous Euclidean description that make use of classical operators.

Based on the above considerations, the total solute concentration ($f$) inside a particle is given by the next equation, which considers the non-adsorbed effective concentration and the amount of solute adsorbed on the surface of porous materials.

$$f = \varepsilon_p C + \rho_p q \qquad (2)$$

where $\varepsilon_p$ and $\rho_p$ are the particle porosity and density, respectively, $C$ is the free solute concentration inside particles, and $q$ is the amount of solute adsorbed, which is assumed to be in equilibrium with $C$.

If one selects an arbitrary fractal region $\Omega_D(t)$ with a volume enclosed by the surface $\partial\Omega_D(t)$, and it assumes that such region contains a constant amount of solute and fluid, the solute mass $\left(m = \int_{\Omega_D(t)} f \, dV_D\right)$ does not change with time, *i.e.*,

$$\frac{d}{dt} \int_{\Omega_D(t)} f \, dV_D = 0 \qquad (3)$$

Here, we have used a fractal measure of the volume of the region, $dV_D$, where $D$ is the mass fractal dimension that considers how the region $\Omega_D(t)$ fills the space where it is embedded and it is independent of the region's shape (Taeasov 2011). From the fractal continuum approach, there are equivalences between the fractal and Euclidean elements (Tarasov 2005, Tarasov 2010, Tarasov 2015, Balankin and Espinoza-Elizarraraz, 2012, and Herrera-Hernández at al., 2013). Such elements map the surface and volume elements of an arbitrary fractal region into a fractal continuum system where the Euclidean description is valid, as is depicted in Figure 1.

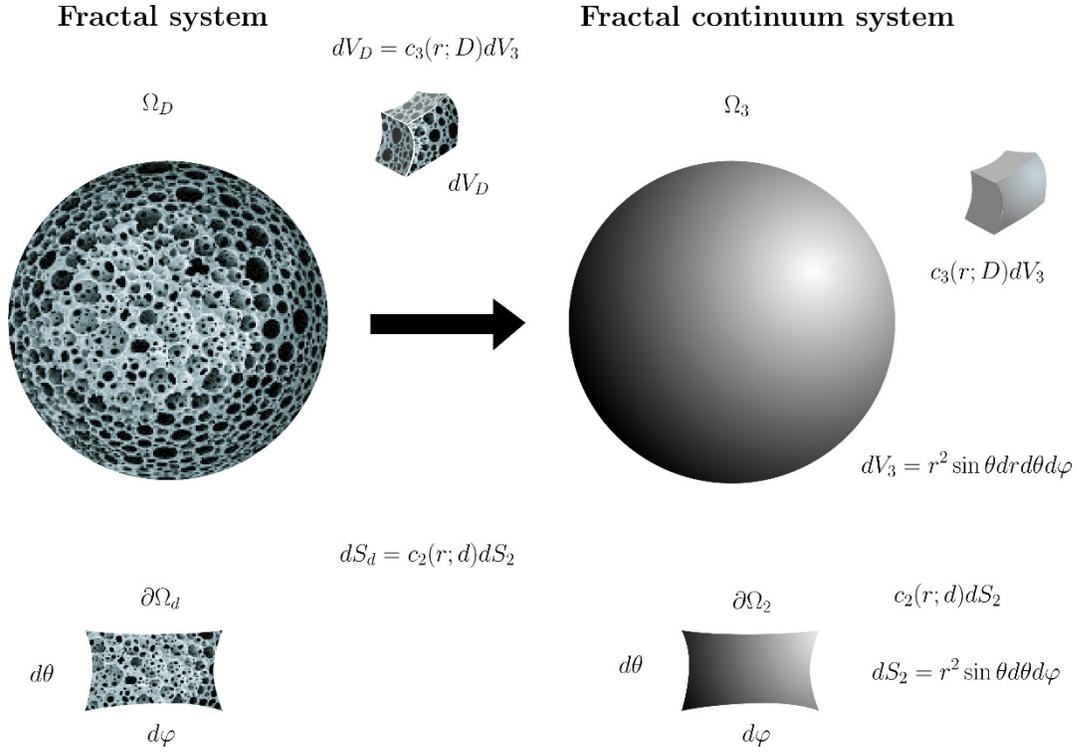

Figure 1. (Color online) Fractal domains (surface and volume) and their mappings to the fractal continuum space through the homogenization coefficients.

In spherical coordinates, the volume element of an arbitrary region has the following correspondence with the Euclidean space:

$$dV_D = c_3(r;D)dV_3 \qquad (4)$$

where

$$c_3(r;D) = \frac{2^{3-D}\,\Gamma(3/2)}{\Gamma(D/2)} r^{D-3}, \qquad 0 < D \leq 3 \qquad (5)$$

Its corresponding surface element in the radial direction is

$$dS_d = c_2(r;d)\,dS_2 \qquad (6)$$

where

$$c_2(r;d) = \frac{2^{2-d}}{\Gamma(d/2)} r^{d-2}, \qquad 1 < d \leq 2 \qquad (7)$$

In the above equations, $dV_D$ and $dS_d$ are the fractal volume and surface elements, respectively, while $dV_3$ and $dS_2$ the corresponding Euclidean ones. In the case of spherical geometry, the Euclidean surface elements are $d\mathbf{S}_2 = d\mathbf{S}_r + d\mathbf{S}_\theta + d\mathbf{S}_\varphi$
($d\mathbf{S}_2 = r^2 \sin\theta\, d\theta\, d\varphi\, \hat{e}_r + r\sin\theta\, dr\, d\varphi\, \hat{e}_\theta + r\, dr\, d\theta\, \hat{e}_\varphi$) whereas the volume element is $dV_3 = r^2 \sin\theta\, dr\, d\theta\, d\varphi$. Since we assume that the process takes place just in the radial direction, we neglect the contributions in the $\theta$ and $\varphi$ directions, and the Euclidean surface and volume elements are reduced to $d\mathbf{S}_2 = d\mathbf{S}_r$ and $dV_3 = 4\pi r^2 dr$, respectively.

After using the Reynolds' transport theorem to obtain the time derivative for the solute mass inside the fractal region, we get

$$\frac{d}{dt}\int_{\Omega_D(t)} f\, dV_D = \int_{\Omega_D(t)} \frac{\partial f}{\partial t} dV_D + \int_{\partial\Omega_d(t)} (\mathbf{J}\cdot\mathbf{n}) dS_d = 0 \qquad (8)$$

Note that this equation considers the fact that the volume and its enclosing surface of an arbitrary region may change as a function of time. Furthermore, it considers only the radial component in the total flux such that the normal unit vector has one component, $\hat{e}_r$.

Using Eq. (6) to change the integration element in the second term of the right-hand side of Eq. (8) leads to

$$\int_{\partial\Omega_d(t)} (\mathbf{J}\cdot\mathbf{n})\, dS_d = \int_{\partial\Omega_2(t)} [c_2(r;d)\mathbf{J}\cdot\mathbf{n}]\, dS_2 \qquad (9)$$

Then, we use the Gauss theorem to change surface integral to a volume integral as

$$\int_{\partial\Omega_2(t)} [c_2(r;d)\mathbf{J}\cdot\mathbf{n}]\, dS_2 = \int_{\Omega_3(t)} \nabla\cdot[c_2(r;d)\mathbf{J}]\, dV_3 \qquad (10)$$

Finally, by using the volume element given in Eq. (4) to change from the Euclidean to the fractal space in Eq. (10), and substituting the resulting equation into Eq. (8), we get

$$\frac{d}{dt}\int_{\Omega_D(t)} f\, dV_D = \int_{\Omega_D(t)} \left\{\frac{\partial f}{\partial t} + \frac{1}{c_3(r;D)}\nabla\cdot[c_2(r;d)\mathbf{J}]\right\} dV_D = 0 \qquad (11)$$

where the only way the right-hand integral is zero is that its argument is zero accordingly. This allows for obtaining a conservation equation for the solute within the particle. I.e.,

$$\frac{\partial f}{\partial t} + \frac{1}{c_3(r;D)}\nabla\cdot[c_2(r;d)\mathbf{J}] = 0 \qquad (12)$$

where **J** is the total solute flux crossing the fractal region through its boundaries, and $c_2(r;d)$ and $c_3(r;D)$ are the surface and volume mapping coefficients (dimensional regularization coefficients) from the fractal space to the Euclidean space, respectively. Such mapping coefficients compensate for the fact that neither the whole surface nor the total volume of a particle is available to the diffusion-adsorption process. The presence of solid material and voids that are randomly distributed, following an invariance principle, restrain the process to a fraction of its surface and volume. Due to the self-similarity in the particles' geometric characteristics, the volume and surface obey a power-law distribution based on their fractal dimensions. For instance, imagine a spherical shell with holes of different sizes distributed randomly, where a solute is allowed to enter into the particle only for some regions of the surface, and by being inside the particle it can be adsorbed only on a fraction of the inner surface of the porous particle. If the surface and volume fractal dimensions remain constant, a dependence on the radial coordinate is required to address the effect of changing the radial position. The mapping coefficients to perform that task are described in Eqs. (4)-(7).

Because fractal dimensions are intrinsic geometrical properties associated with the adsorbent particles, they do not depend on the solute shape or size if and only if the average pore size is much larger than the solute size. In this way, it is possible to determine, by matching with the experimental data, such geometric characteristics for a given adsorbent material by using different solutes.

The solute mass flux has two contributions: one is due to the diffusion process taking place inside the adsorbent particle where the driving force is the free solute concentration, and the other occurs on the porous surface and has as a driving force on the solute gradients on the porous surface at the porous network. Both contributions are affected by the surface mapping coefficient according to the expression

$$c_2(r;d)\mathbf{J} = c_2(r;d)\left( \underbrace{\mathbf{J}_p}_{Intra-particle} + \overset{Surface}{\overbrace{\mathbf{J}_s}} \right) \quad (13)$$

where the intraparticle flux is given by

$$\mathbf{J}_p = -D_p \nabla C \quad (14)$$

and the superficial flux is represented by the following equation

$$\mathbf{J}_s = -D_s \rho_p \nabla q \quad (15)$$

Note that even though we are considering a fractal particle, in this model's formulation, we assume that the diffusion process takes a regular form, and so we use the classic Fickian-like expressions for mass fluxes. Modifications to the flux to incorporate the space/time memory effects that the solute exhibits while moving within adsorbent spheres are the main topic of one ongoing work. Based on Eq. (2), the accumulation term can be rewritten as follows:

$$\frac{\partial f}{\partial t} = \varepsilon_p \frac{\partial C}{\partial t} + \rho_p \frac{\partial q}{\partial t} \tag{16}$$

By substituting Eqs. (14) and (15) into Eq. (13), and the resulting expression and (16) into Eq. (12), we get

$$\varepsilon_p \frac{\partial C}{\partial t} + \rho_p \frac{\partial q}{\partial t} - \left(\frac{\hat{c}_2}{\hat{c}_3}\right) r^{1-D} \frac{\partial}{\partial r}\left( D_p\, r^d \frac{\partial C}{\partial r} + D_s \rho_p r^d \frac{\partial q}{\partial r} \right) = 0 \tag{17}$$

where $\hat{c}_2/\hat{c}_3 = \dfrac{2^{2-d}}{\Gamma(d/2)} \Big/ \dfrac{2^{3-D}\,\Gamma(3/2)}{\Gamma(D/2)}$ is the relationship between the constant parts of the surface and volume scaling that depend on their fractal dimensions.

At this point, we need a relationship between the free concentration inside the adsorbent particles and the amount of solute adsorbed on the active sites. It is common to propose an empiric equilibrium relationship like

$$q = F\left[C(r,t)\right] \tag{18}$$

From the above relation, we can get $\partial q/\partial t$ and $\partial q/\partial r$, which complement Eq. (17) as follows:

$$\frac{\partial q}{\partial t} = \left(\frac{\partial q}{\partial C}\right)_r \frac{\partial C}{\partial t} \tag{19}$$

$$\frac{\partial q}{\partial r} = \left(\frac{\partial q}{\partial C}\right)_t \frac{\partial C}{\partial r} \tag{20}$$

By using Eqs. (19) and (20) and doing some algebra, the mathematical model given in (17) is reduced to

$$\frac{\partial C}{\partial t} - \frac{1}{\zeta(r,t)} \frac{\partial}{\partial r}\left[\psi(r,t) \frac{\partial C}{\partial r}\right] = 0 \tag{21}$$

where the space-time-dependent coefficients $\zeta(r,t)$ and $\psi(r,t)$ are given by

$$\zeta(r,t) = \left(\varepsilon_p + \rho_p \frac{\partial F}{\partial C}\right) \hat{c}_3 r^{D-1} \tag{22}$$

and

$$\psi(r,t) = \left(D_p + D_s \rho_p \frac{\partial F}{\partial C}\right) \hat{c}_2 r^d \tag{23}$$

Finally, the mathematical model for the intraparticle adsorption-diffusion is described by the following nonlinear second-order partial differential equation:

$$\frac{\partial C}{\partial t} = \frac{1}{\zeta(r,t)} \frac{\partial \psi(r,t)}{\partial r} \frac{\partial C}{\partial r} + \frac{\psi(r,t)}{\zeta(r,t)} \frac{\partial^2 C}{\partial r^2} \tag{24}$$

which is a generalization of the adsorption-diffusion model in the spherical coordinates for any equilibrium relationship. In the Euclidean limit case when $d = 2$ and $D = 3$, the classical adsorption-diffusion model is recovered. The analytical solution to Eq. (24) is not available and it is not possible to obtain since it strongly depends on the equilibrium relationship, which most of the time is a nonlinear function. Instead of trying the analytical way, in the next section, we explore a numerical alternative through the method of lines.

## 3. Numerical solution: Method of lines

To solve the mathematical model given in Eq. (24), we discretize the first and second order space partial derivatives with symmetric finite difference approaches, thereby leaving the time partial derivative continuous. It is known as the method of lines. The discretization of the spatial operators through the finite differences method (or finite element method) results in a first-order differential equations system. The space domain $r \in (0, R_p]$ discretization is carried out by defining the node number ($M$) so that for a uniform radial mesh we have $\Delta r = R_p / M$, $r_i = r_{i-1} + (i-1)\Delta r$ for $2 \leq i \leq M$ and $r_1 = \Delta r$. To deal with the singularity at the origin, we replace such a position with one radial increment and the number of spatial nodes was $M = 1000$. Each discretization element $r_i$ corresponds to a sphere whose governing equation is given by

$$\frac{dC_i}{dt} = \frac{1}{\zeta_i(t)} \frac{\psi_{i+1}(t) - \psi_{i-1}(t)}{2\Delta r} \frac{C_{i+1}(t) - C_{i-1}(t)}{2\Delta r} + \frac{\psi_i(t)}{\zeta_i(t)} \frac{C_{i-1}(t) - 2C_i(t) + C_{i+1}(t)}{\Delta r^2} \tag{25}$$

which has an equivalent representation as

$$\begin{aligned}\frac{dC_i}{dt} &= \left[\alpha_i(t)\psi_{i-1}(t) + \beta_i(t) - \alpha_i(t)\psi_{i+1}(t)\right]C_{i-1}(t) - 2\beta_i(t)C_i(t) \\ &+ \left[\alpha_i(t)\psi_{i+1}(t) - \alpha_i(t)\psi_{i-1}(t) + \beta_i(t)\right]C_{i+1}(t)\end{aligned}, \quad 2 \leq i \leq M-1$$

(26)

Here, coefficients $\alpha_i(t)$ and $\beta_i(t)$ have the following expressions:

$$\alpha_i(t) = \frac{1}{4\Delta r^2 \zeta_i(t)} \qquad (27)$$

$$\beta_i(t) = \frac{\psi_i(t)}{\Delta r^2 \zeta_i(t)} \qquad (28)$$

For the first and the last nodes, a special treatment is necessary in which the first node zero-flux condition is imposed and in the last node the continuity of the mass flux is set. Then,

$$-\mathbf{n}_{p-s} \cdot \left(D_p \nabla C + D_s \rho_p \nabla q\right) = k_L \left(C_A - C\big|_{B_{p-s}}\right), \quad \text{at} \quad i = 1 \qquad (29)$$

From this equation, we get the ghost node appearing in (26) at the boundary when $i = M$.

$$C_{M+1}(t) = C_{M-1}(t) + \frac{k(C_A(t) - C_M(t))}{\dfrac{D_p}{2\Delta r} + \dfrac{D_s \rho_p}{2\Delta r} \dfrac{\partial q}{\partial C(t)}\bigg|_{i=M}} \qquad (30)$$

At the center of the adsorbent particle, we set

$$\frac{\partial C(t)}{\partial r}\bigg|_{r=0} = 0 \qquad (31)$$

From this, we get the ghost node of in Eq. (26) for $i = 0$.

$$C_0(t) = C_2(t) \qquad (32)$$

## 4. Intraparticle mass transport

The solution to Eq. (24) gives the spatial distribution of the free solute concentration inside an adsorbent particle at any time interval, $C(r,t)$. It is well-known that in Euclidean systems the first and second moments are given by $\langle r(t) \rangle \sim t^{1/2}$ and $\langle r(t)^2 \rangle \sim t$, respectively. Here, the proportional constants depend on the embedded Euclidean dimension and the diffusion coefficient. In this sense, the Mean Square Displacement (MSD) scales as $\sigma^2(t) = \langle (r(t) - \langle r(t) \rangle)^2 \rangle \sim Dt$. However, considering the fractality (and high-tortuous paths, channels, and porous surface) in the adsorbent particles, a more complex dependence and behavior of $\sigma^2(t)$ is expected. Having in mind such arguments, and considering the fact that at the beginning of the adsorption-diffusion process the solute concentration in the solution is higher than the solute concentration inside adsorbent particles, the concentration gradients point toward the center of these. Hence, it is possible to calculate the *n-th* moment of the concentration distribution as

$$\left\langle r(t)^n \right\rangle = \frac{\int_R^0 C(R-r,t)(R-r)^{D+n-1} dr}{\int_R^0 C(R-r,t)(R-r)^{D-1} dr} \tag{33}$$

Notice that the integration interval is from the solution (particle surface) to the center of the adsorbent particles, which is a consequence of the concentration gradients' orientations. The first moment, $n = 1$, is calculated according to

$$\left\langle r(t) \right\rangle = \frac{\int_R^0 C(R-r,t)(R-r)^{D} dr}{\int_R^0 C(R-r,t)(R-r)^{D-1} dr} \tag{34}$$

The second moment, $n = 2$, is

$$\left\langle r(t)^2 \right\rangle = \frac{\int_R^0 C(R-r,t)(R-r)^{D+1} dr}{\int_R^0 C(R-r,t)(R-r)^{D-1} dr} \tag{35}$$

With the first two moments, the MSD is evaluated according to

$$\sigma^2(t) = \left\langle \left[ r(t) - \left\langle r(t) \right\rangle \right]^2 \right\rangle = \left\langle r(t)^2 \right\rangle - \left\langle r(t) \right\rangle^2 \tag{36}$$

whose explicit form, which is obtained by substituting Eqs. (34) and (35) into Eq. (36), is

$$\sigma^2(t) = \frac{\int_R^0 C(R-r,t)(R-r)^{D+1} dr}{\int_R^0 C(R-r,t)(R-r)^{D-1} dr} - \left( \frac{\int_R^0 C(R-r,t)(R-r)^{D} dr}{\int_R^0 C(R-r,t)(R-r)^{D-1} dr} \right)^2 \tag{37}$$

In particle dynamics, the MSD is associated with the dispersion phenomenon, which is a measurement of the particle distribution within a specific domain. In our case, this time-dependent property describes the solute distribution inside the spherical particle. Since it is time-dependent, MSD allows us to know how the solute saturates the porous media while the adsorption-diffusion phenomenon takes place. In an ongoing work, we relate the MSD with transport behavior in order to estimate the normal, subdiffusion and superdiffusion regimes.

## 5. Estimation of fractal dimensions: Adsorption on activated carbon

In this section, we present the application of the fractal-diffusion mathematical model to simulate the adsorption of solutes onto one adsorbent material. It is done in order to demonstrate the applicability and accuracy of the mathematical model to reproduce real data, and with this aim we will model the experimental adsorption of three substances onto granular activated carbon (GAC). The adsorbed substances are phenol, Methylene Blue (MB) and Methyl Blue (MTB). The systems' phenol-GAC, MB-GAC, and MTB-GAC have been the subject of study in a previous work (Ocampo-Perez et al., 2013). Table 1 presents the most relevant information from the experimental point of view and the equilibrium constants. The adsorption equilibrium was obtained using a batch adsorber at pH 3 and T=298 K. The experimental adsorption equilibrium data of phenol, MB, and MTB were interpreted by using the Langmuir adsorption isotherm model given by (39) whose parameters are in the last two columns of Table 1.

$$q(C) = \frac{q_m K C}{1 + K C} \qquad (39)$$

**Table 1**. Physicochemical properties of adsorbates and Langmuir adsorption isotherm constants

| Adsorbate | Molecular structure | Molecular Formula | $pK_a$ | Dimensions X, Y, Z (nm) | $q_m$ (mg/g) | K (L/mg) |
|---|---|---|---|---|---|---|
| Phenol | 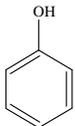 | $C_6H_6O$ | 9.86 | 0.675 0.706 0.296 | 182.9 | 0.0189 |
| MB | 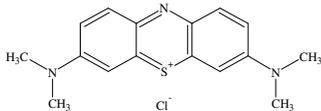 | $C_{16}H_{18}N_3SCl$ | 3.80 | 1.641 0.744 0.617 | 200.7 | 0.0156 |

| | | | | | 2.1 | |
| --- | --- | --- | --- | --- | --- | --- |
| MTB | 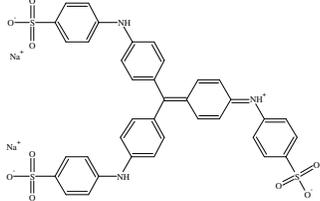 | $C_{37}H_{27}N_3Na_2O_9S_3$ | 8.80 | 1.36 | 63.77 | 0.027 |
| | | | | 0.81 | | |

Figure 2 shows the simulation of decay curves for the adsorption of the three substances [a) for phenol, b) for MB, and c) for MTB]. The simulations are referred to as the numerical results of the fractal-diffusion model using the method of the lines-scheme presented in a previous section. In general, an excellent agreement is obtained except for the MTB-GAC system where the equilibrium conditions were not reached. The matching between the experimental and simulated curves was carried out by the numerical optimization of the model parameters $D$ and $d$. The optimization of these parameters was subjected to the minimization of the following error:

$$\text{error}(\mathbf{p}) = \int_0^{t_f} \left| C(t;\mathbf{p}) - C^{\exp}(t) \right| dt \qquad (40)$$

where $C(t;\mathbf{p})$ is the solution concentration from the numerical solution of the mathematical model, $C^{\exp}(t)$ is the laboratory solution concentration, and $\mathbf{p} = \{d, D\}$ is a vector of the model parameters. The optimum values of the fractal dimensions and the minimum computed error are summarized in Table 2, while the error surfaces are plotted as subfigures within the decay curves in Figure 2.

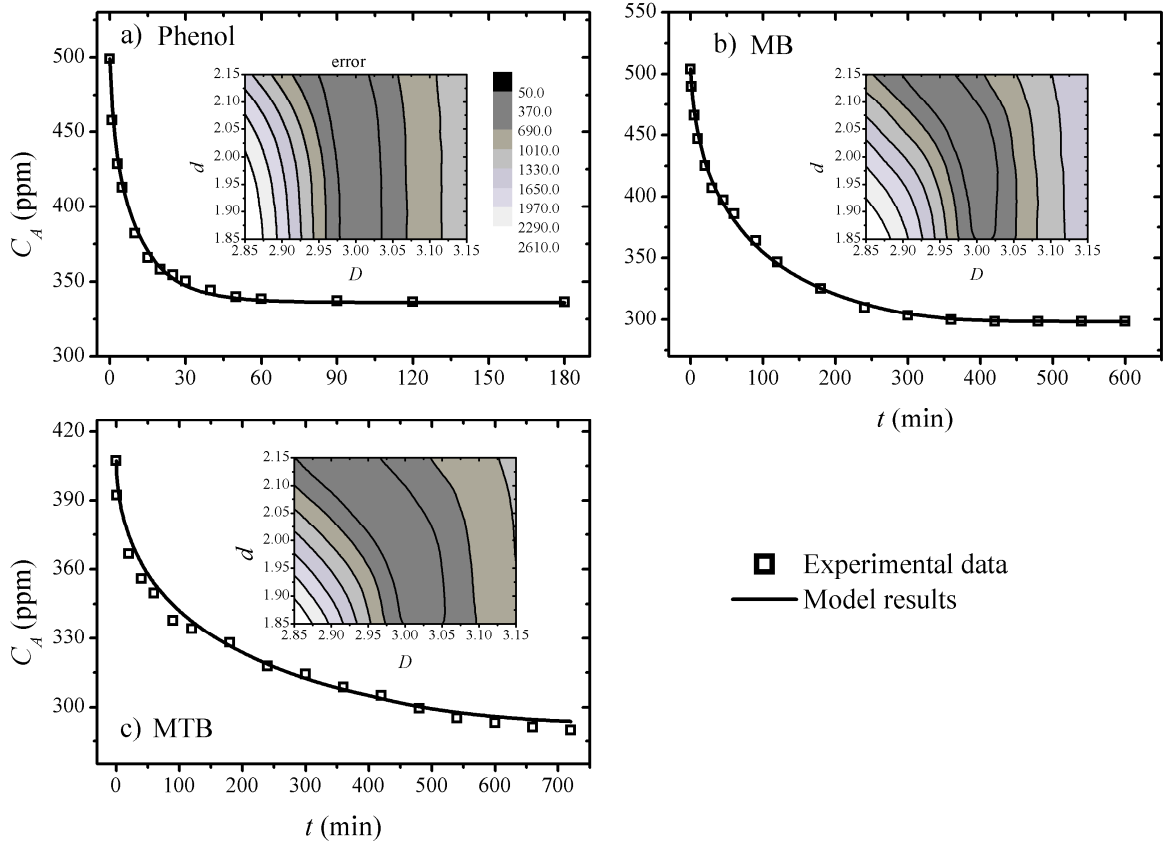

**Figure 2**. (Color online) Decay curves for the adsorption of three substances onto activated carbon: a) phenol, b) MB and c) MTB. Here, the numerical results of the fractal-adsorption model were fitted to experimental data and the error surface [ppm·s] is plotted as a function of the fractal parameters ($D$ and $d$).

**Table 2**. Model parameters for the adsorption of the three substances onto activated carbon.

| Parameter | Phenol | MB | MTB |
|---|---|---|---|
| $D$ | 3.000 | 2.997 | 3.000 |
| $d$ | 1.991 | 2.010 | 1.997 |
| *Error, ppm s | 13.093 | 48.391 | 98.054 |

*Error computed between experimental data and model results.

It was found that for GAC, the fractal dimension $D = [2.997 - 3.000]$ while $d = [1.991 - 2.010]$. I.e., the fractal geometries for activated carbon are close to the Euclidian ones. As mentioned previously, the fractal dimensions weakly depend on the adsorbed solute. The surface fractal dimension has been reported previously by Cuerda-Correa et al. (2006) and Wang et al. (2007). Cuerda-Correa et al. determine this fractal property for activated carbon (coconut, Aldrich and Merck) within the interval of $2.81 \leq d \leq 2.85$ using a thermogravimetric technique and

$2.77 \leq d \leq 2.85$ using $N_2$ adsorption. While Wang et al. have estimated that $0.99 \leq d \leq 1.58$ for GAC-ZJ15, they found that it depends on the adsorbed substance and the initial concentration. Significant variations are noted among the reported values and our estimations. This presumably is attributed to the theoretical background behind the fractal models used in each case. That is, previous works based their observations on a fractal model for adsorption isotherms, while in our work, a new geometrical fractal adsorption-diffusion model is proposed, and we must recall that our theory does not allow for a solute-dependent fractal dimension, as was verified in our results presented in Table 2.

The intraparticle concentration profiles for the three adsorbed substances are graphed in Figures 3a), b) and c), where each curve corresponds to a given time. Note that the time to reach equilibrium conditions is different for each solute, thus indicating that the adsorption capacity and kinetics of GAC depend on the adsorbed molecule. The tendency of concentration profiles takes place following the expected behavior. I.e., for each time, the concentration is larger the closer it is to the particle's surface. This can be thought of as one normal mass transport process inside the GAC where the mass flux vector is always directed toward the center of particle; however, in our numerical simulations, one particular and contradictory case was found. The concentration profiles for the MB of Figure 3b show that for $5.6\,\text{h} < t < 7\,\text{h}$, the mass flux in the center of the particle is directed toward the surface particle, thereby indicating one coupled phenomenon of adsorption-desorption in that zone. So far, to this point, the results plotted in Figures 2 and 3 demonstrate that the adsorption-diffusion-fractal model is able to accurately predict real lab experiments, and to calculate the intraparticle concentration.

In this way and taking as a benchmark case the fitted model for the adsorption of MB, in the following section, we will investigate the effects of fractal parameters $D$ and $d$ over decay curves and intraparticle concentration profiles.

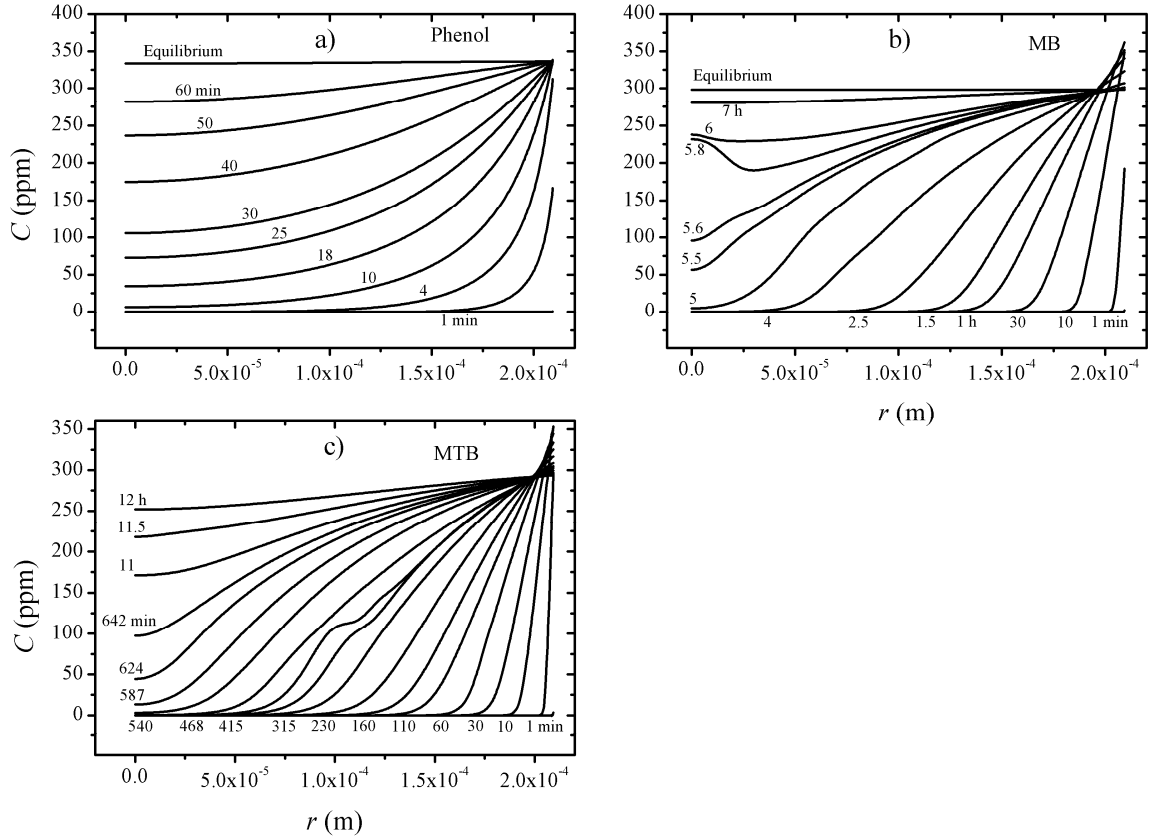

**Figure 3**. (Color online) Evolution of intraparticle solute concentration profiles for three substances adsorbed onto activated carbon: a) phenol, b) MB and c) MTB.

On one hand, besides the solution and intraparticle concentrations presented in previous figures, the total radial mass flux inside the adsorbent particle was evaluated at a given position. The middle $r = R/2$ was chosen with the aim to avoid the boundary effects from the GAC surface and meet the symmetry conditions at the center of the adsorbent, $r = 0$. The mass fluxes for the three adsorbent systems phenol-GAC, MB-GAC and MTB-GAC are plotted in Figures 4a), b) and c), respectively, for the optimal parameter values. We must recall that the total radial mass flux was calculated according to Eq. (13) where one coupled interdependency between the pore-volume and surface effective diffusion is considered. The contribution of each type of diffusion is not presented since it is beyond the scope of this work, but it is mentioned that the mass transport of some regions inside the GAC can be governed by one type of diffusion at certain times. This phenomenon has been reported in the literature in the study of the adsorption of acetaminophen onto GAC by Ocampo-Pérez et al. (2017), and some extrapolations to our cases are direct. Now, turning our attention to results plotted in Figure 4, it is noticed that for all the cases three general regimes can be defined:

i) Increasing the mass flux at the beginning of adsorption,

ii) The maximum mass flux has been reached, and

iii) One eventual mass flux decreasing regime.

The fluctuations of the mass flux can be attributed to the complex interrelation between the pore-volume, surface diffusion and adsorption-desorption of the solute. The rate of each one of these phenomena is not constant with time, and individual variations cause the driving force of each process to also vary with time.

On the other hand, the increasing mass flux regime observed at early times for the results plotted in Figure 4 indicates that optimum conditions are met for the maximum transport of solute from the solution toward the interior of the GAC. For instance, one of the optimum conditions is the fact that there is no solute inside the pores or over the surface of the GAC. This maximizes the driving forces ($\nabla C$ for pore-volume diffusion, $\nabla q$ for surface diffusion, and $C - C^{equilibrium}$ for adsorption-desorption) for transport and adsorption. In addition, the *strength* of diffusion is analyzed in terms of the MSD presented in Eq. (37), and for this purpose we have also plotted $\sigma^2$ for all the adsorbent systems in Figure 5. We note that the transport behavior is different for each adsorbate. It depends on the size-dependent diffusion coefficient. For phenol molecules, its MSD changes abruptly in the short-term, and its maximum value is significantly higher than for the other cases. This result suggests that solute enters into the adsorbent particles with almost no restrictions, and it uniformly covers its surface. The other solutes are much larger than phenol (size of MTB >> MB >> phenol, as indicated in the schemes depicted in Figure 4), and this directly impact the mass flux and the equilibrium, as seen in Figures 4 and 5. The maximum value of the mass flux agrees well with the minimum value of the solute size. For phenol, it is greater than $2.0 \times 10^{-2}$ (in SI units), for MB it is around $4.0 \times 10^{-3}$ and for MTB it is about $6.0 \times 10^{-4}$. The same behavior is observed in the MSD trends, which is an indicator of the solute dispersion within adsorbent particles, where it is inversely proportional to the solute size, and the smallest the solute size the greatest the dispersion. Another fact observed in Figure 5 is that before reaching the maximum MSD value, various slope changes are present for MB and MTB. Such changes are a consequence of the solute size and shape, and the way that the solutes interact with the GAC.

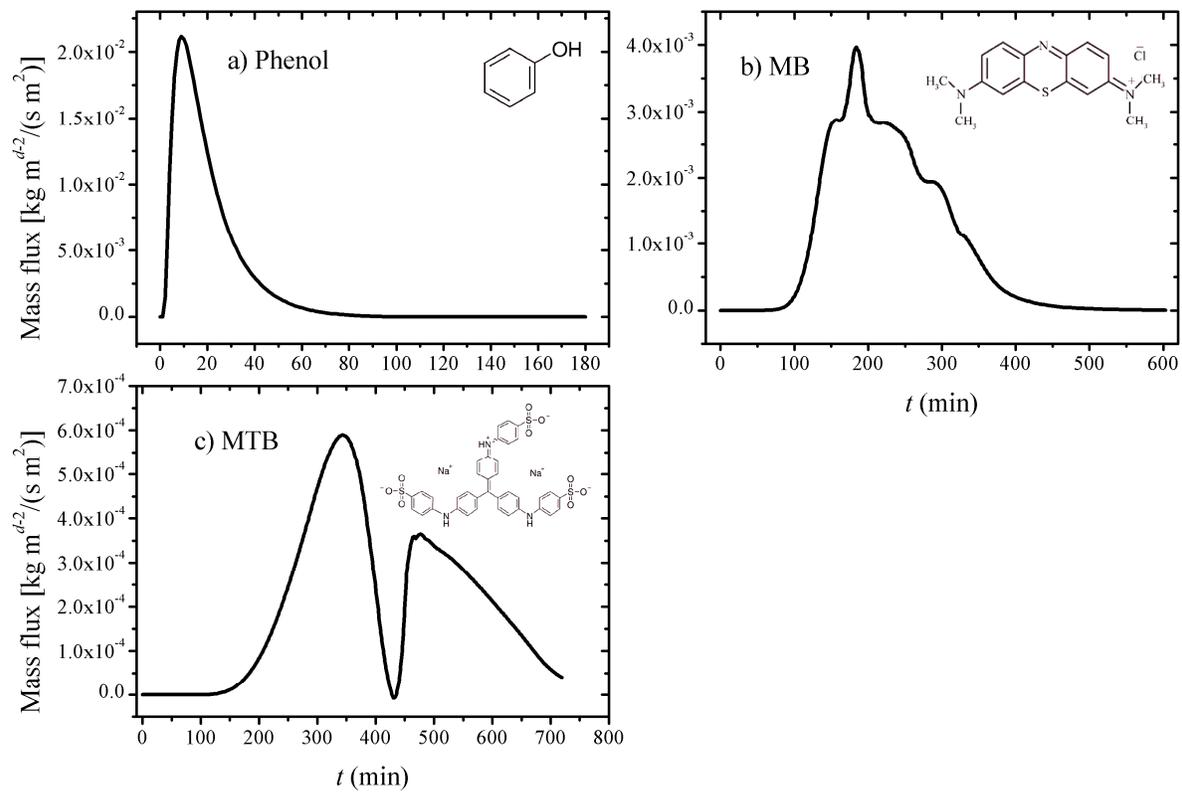

**Figure 4**. (Color online) Evolution of total radial mass flux evaluated at $r = R/2$ for the adsorption onto GAC of three substances: a) phenol, b) MB, and c) MTB and for the optimal fractal parameters.

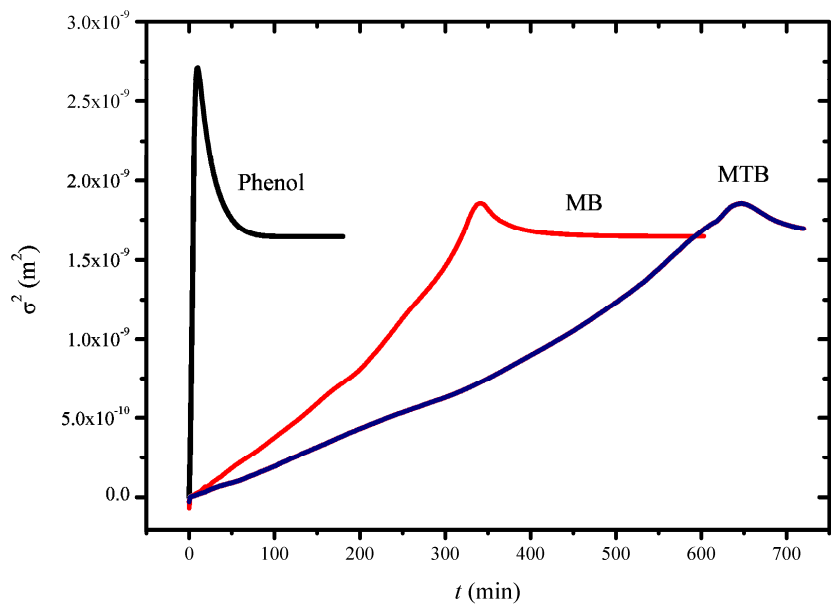

**Figure 5**. Evolution of MSD for adsorption of a) phenol, b) MB and c) MTB onto GAC.

## 6. Numerical analysis of fractal dimensions

On the basis of the fitted mathematical model of the lab experiments of MB onto GAC, we conducted a sensitivity analysis of the fractal dimensions of surface and volume, $d$ and $D$, respectively, with the aim to assess the impacts of changes over the behavior of the adsorption process. With this purpose, Figure 6 presents the solution and intraparticle concentrations profiles when such parameters were varied to be $D = [2.9 - 3.05]$ and $d = [1.8 - 2.2]$ (we recall that while the maximum value of $D$ is bounded to 3, nevertheless here we included simulations up to $D = 3.05$ as an illustration of its impact over the adsorption process). In general, it is shown that the volume fractal dimension $D$ changes drastically the dynamics of the adsorption process and the time at which the equilibrium is reached. In the case of the solution concentration profile, as shown in Figure 6a, it is apparent that lower values of $D$ result in an increment in the transitory stage, and so achieving the equilibrium takes longer. Moreover, the equilibrium concentration value is reduced according to the reduction in the fractal dimension. Regarding the intraparticle concentration, as shown in Figure 6b, the radial profile strongly depends on $D$, and a slight reduction of $D$ with respect to the Euclidean benchmark case ($D = 3$) implies a radial concentration profile modification. It is interesting to compare the cases where $D = 2.9$ and $D = 2.95$ with the optimal case $D = 2.9973$. In the first two cases, the solute concentration is zero for a radial position smaller than $5.5 \times 10^{-5}$ m, whereas for the optimal case we observe a decreasing profile toward the origin without reaching a zero value at this point.

The surface fractal dimension effect on the solution concentration decay is shown in Figure 6c, while the intraparticle radial concentration profile is presented in Figure 6d. As in the case of the volume fractal dimension, this parameter sensibly affects not only the concentration dynamics but also the intraparticle concentration profiles. When analyzing the solute concentration in the solution, as shown in Figure 4c, we observe that a decrement in $d$ slightly affects the concentration equilibrium value if $d \leq 2$ and the adsorption dynamics are faster than for $d > 2$. This implies that the time necessary to reach the equilibrium point decreases as $d$ decreases. For example, for $d = 1.8$, the equilibrium time is approximately 75 *min.*, whereas for the optimal case ($d = 2.0103$) it is approximately 400 *min*. On the other hand, the effect of $d$ on the intraparticle radial profile is opposite to the behavior described by $D$ since the values of $d \leq 2$ result in a homogenous concentration inside the particle, whereas the same behavior is also observed for $D > 3$.

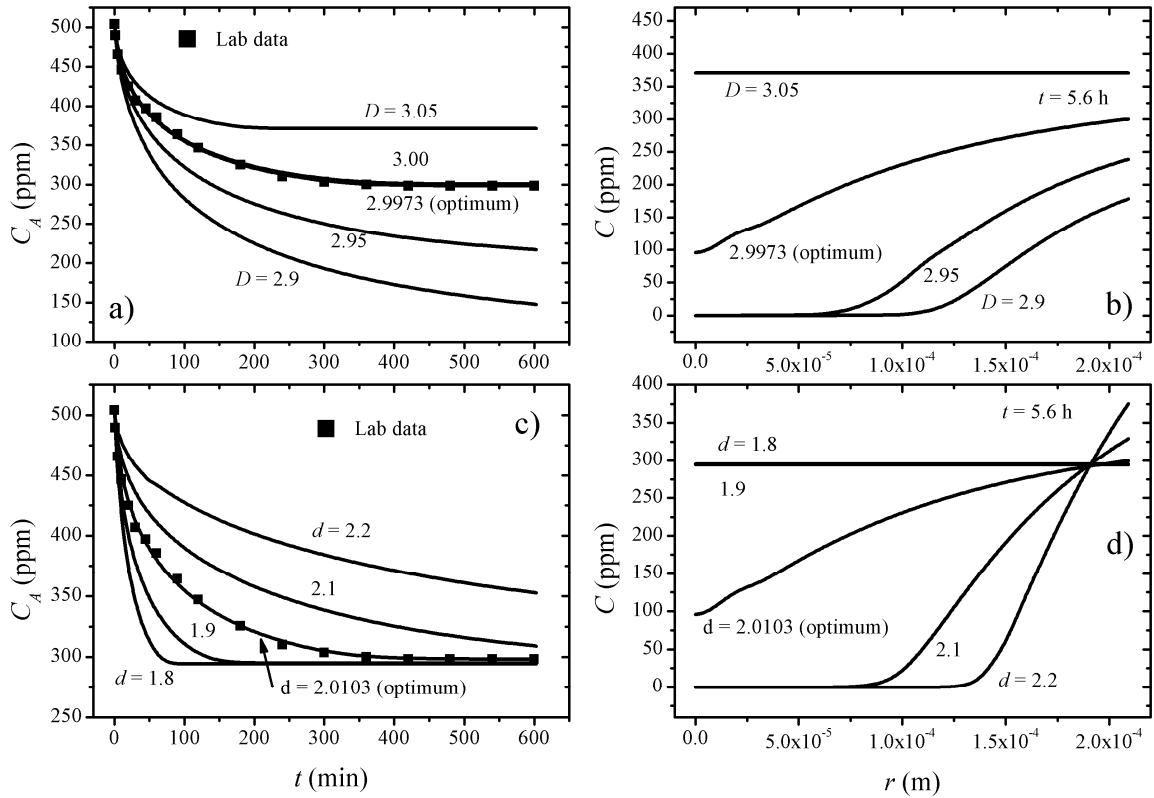

**Figure 6**. (Color online) Decay curves [a) and b)] and intraparticle concentration profiles [b) and d)] for the adsorption of MB onto GAC. Here, the dimensional parameters $D$ and $d$ were varied, and for the intraparticle concentrations, the time was fixed to be $t = 5.6$ h.

The impacts of the fractal dimensions over the total radial mass flux crossing a spherical shell located at $r = R/2$ are presented in Figure 7. There, we have plotted the variations of the mass flux as a function of the time and fractal dimensions of volume $D$ in Figure 7a and surface $d$ in Figure 7b. It was found that both fractal parameters oppositely affect the radial flux. It is evident that such fractal dimensions affect not only the maximum value of the mass flux but also the time interval at which the mass flux is different from zero. If the optimal case (black line) is taken as the benchmark case in Figures 7a and 7b, one finds that an increment in $D$ (decrease in $d$) is translated into an increase (reduction) of the maximum value of the flux and a concurrent decrease (increment) in the time interval at which the mass flux remains constant with zero value. Note that it is possible to evaluate the amount of mass crossing the hypothetical spherical shell by integrating the total mass flux over some time interval of interest. Such a quantity is proportional to the curve width and the curve height. For instance, it decreases as $D$ increases regarding the benchmark case. The opposite case can be inferred from the flux behavior while changing $d$.

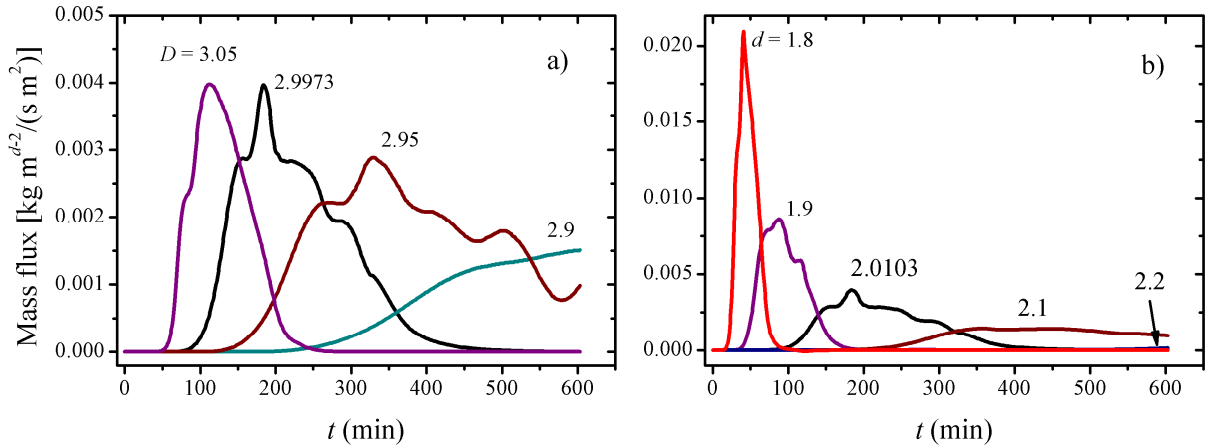

**Figure 7**. (Color online) Variations of the mass flux evaluated at $r = R/2$ for the adsorption of MTB onto GAC as a function of time and the dimensional parameters: a) $D$ and b) $d$. We set the optimal value of $d$ in the first case and the optimal value of $D$ in the second one.

Finally, we present the MSD as a function of time and vary the fractal dimension parameters with respect to adsorption of MB onto GAC, as shown in Figure 8. It sensibly depends on the fractal dimensions, and for most cases, different diffusive regimes (changes in the slope) are observed, but the appearance of them depends on time. Taking the optimum case as the benchmark, larger values of $D$ (or lower values of $d$) reduce (increase) the time at which the MSD stops changing and takes a constant value. This condition is related to the equilibrium process and is verified by the time when the MSD is constant. Moreover, if one compares Figures 6a and 6c with Figures 8a and 8b, one finds that the time at which the concentration decay in the solution and the MSD is constant is the same in both cases. This implies that local gradients inside and outside the adsorbent particles disappear, and a constant MSD could be a consequence of the adsorption-desorption process taking place at the same ratio. The long-term behavior of the MSD is not the same for $d$ and $D$. In the former case, the asymptotic value is the same for all $d$ s, while in the latter case it changes according to the value of $D$.

Since the amount of mass enclosed by a certain fractal volume satisfies the power-law $M \sim r^D$ for some scalar of constant density, it is apparent that greater values of $D$ tend to neutralize the local gradients inside the adsorbent particles so that the MSD gets at a constant value, as shown in Figure 8a. On the other hand, $d$ controls the available surface area that solute particles cross before being adsorbed. It is clear that the smaller that this value is, the lesser the available area is, and the shorter distance that the path particles have to traverse before getting trapped. For example, imagine a particle moving in a highly tortuous two-dimensional path (like Koch curves). If one evaluates the trajectory length traced by such a particle, one would find that it corresponds to a surface with a fractal dimension greater than two, which would be in agreement with the fact that the MSD can be less than those determined for $d < 2$, as is presented in Figure 8b.

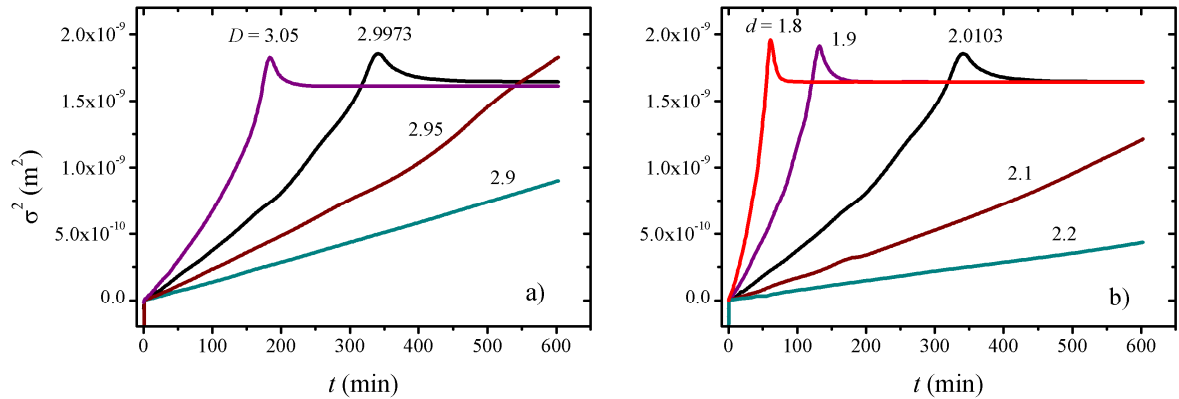

**Figure 8**. (Color online) Evolution of MSD for the adsorption of MB onto GAC and as a function of fractal dimensions of volume ($D$) and surface ($d$). We set the optimal value of $d$ in the first case and the optimal value of $D$ in the second one.

## 7. Concluding remarks

We have presented one mathematical model for the adsorption, pore-volume and surface diffusion of solutes inside one adsorbent and porous material with a fractal microstructure. This model is based on the fractal continuum approach and contemplates the scale-invariant property of the surface and volume of adsorbent particles, which are assumed to be fractals whose complete characterization is made by their corresponding fractal dimensions. The fractal adsorption-diffusion model was applied to simulate and predict the adsorption of three substances onto granular activated carbon. The numerical predictions of the decay curves agree well with the lab data, and subsequently, one of these validated cases was used to investigate the effects of fractal dimensions over relevant variables (the solution and intraparticle concentrations, mass flux crossing a specific area, and mean square displacement). The effect of fractal dimensions on the adsorption-diffusion process is apparent if the optimal case is taken as the benchmark. Moreover, an increment in $D$ (reduction of $d$) provokes a reduction (increment) of the time at which the equilibrium is achieved, which is reflected on the mass flux and also on the MSD. The fractal model demonstrated to be able to predict adsorption experiments, and jointly can be used to estimate the fractal parameters of porous adsorbents. In our case, we found that the fractal dimensions of granular activated carbon are close to the Euclidian ones, and they are almost independent of the adsorbed solute.

**Acknowledgment**

ECH-H acknowledges the support from CONACYT trough cátedra at CIDESI. MN-L acknowledges the financial support from the Asociación Mexicana de Cultura, A.C.